\documentclass[conference]{IEEEtran}
\IEEEoverridecommandlockouts
\usepackage{cite}
\usepackage{amsmath,amssymb,amsfonts}
\usepackage{graphicx}
\usepackage{textcomp}
\usepackage{xcolor}
\usepackage{booktabs}
\usepackage{tikz}
\usepackage{pgfplots}
\usepackage{caption}
\usepackage{subcaption}
\usepackage{url}
\usetikzlibrary{arrows.meta, positioning, shapes.geometric, fit, calc, backgrounds}
\pgfplotsset{compat=1.18}

\def\BibTeX{{\rm B\kern-.05em{\sc i\kern-.025em b}\kern-.08em
    T\kern-.1667em\lower.7ex\hbox{E}\kern-.125emX}}

\begin{document}
\linespread{0.97}\selectfont

\title{Cooperative Platoon Routing and Dispatching via Edge-Assisted Hybrid Quantum Optimization}

\author{\IEEEauthorblockN{Talha Azfar}
\IEEEauthorblockA{\textit{Department of Civil and Environmental Engineering} \\
\textit{Rensselaer Polytechnic Institute}\\
Troy, NY, USA \\
azfart@rpi.edu}
\and
\IEEEauthorblockN{Ruimin Ke}
\IEEEauthorblockA{\textit{Department of Civil and Environmental Engineering} \\
\textit{Rensselaer Polytechnic Institute}\\
Troy, NY, USA \\
ker@rpi.edu}
}

% \author{\IEEEauthorblockN{-- --}
% \IEEEauthorblockA{\textit{-- -- --} \\
% \textit{-- -- --}\\
% -- -- -- \\
% --}
% \and
% \IEEEauthorblockN{-- --}
% \IEEEauthorblockA{\textit{-- -- --} \\
% \textit{-- -- --}\\
% -- -- -- \\
% --}
% }

\maketitle
\thispagestyle{plain}
\pagestyle{plain}

\begin{abstract}
Cooperative platooning can reduce the energy use of Connected and Autonomous Vehicle (CAV) fleets, but the routing problem becomes difficult when vehicles must meet on the same road segments at compatible times while moving through unstable urban traffic. This paper develops an edge-assisted, closed-loop evaluation pipeline for platooning-aware vehicle routing. Roadside Units estimate local traffic kinematics from video, classify segment-level flow stability, and activate platooning rewards only on road segments where close-gap coordination is physically appropriate. The resulting multi-vehicle routing problem is written directly as a Quadratic Unconstrained Binary Optimization (QUBO) model, so pairwise platooning interactions are represented as native quadratic Ising terms instead of requiring auxiliary MILP linearization variables. We evaluate the framework using a 24-hour microscopic SUMO simulation of Troy, NY, together with localized IBM Quantum hardware benchmarks. The SUMO study shows an 18.5\% reduction in fleet tractive-energy demand relative to a non-cooperative baseline. On 25-active-qubit benchmark instances executed on \texttt{ibm\_boston}, Linear-Chain QAOA reduces two-qubit CNOT depth by 66.7\% compared with dense QAOA and samples the exact classical ground state with $P_{\text{feas}} = 38.6\%$ and $P_{\text{opt}} = 14.2\%$ at $p=2$. These results suggest that edge perception and shallow quantum optimization can work together as a useful component of closed-loop CAV platoon dispatching.
\end{abstract}


\section{Introduction}
\label{sec:intro}

Connected and Autonomous Vehicles (CAVs) enable cooperative fleet management through vehicle platooning, where CAVs travel in close proximity at synchronized velocities to reduce aerodynamic drag and energy consumption~\cite{lammert2014effect, alam2010experimental}. However, dynamically routing fleets to maximize platooning opportunities across urban networks remains challenging, as vehicles must synchronize both spatial paths and departure schedules.

Classical Mixed-Integer Linear Programming (MILP) models for Vehicle Routing with Cooperative Platooning (VRP-CP) introduce continuous arrival times and auxiliary binary variables to linearize non-convex quadratic V2V coupling terms. However, these auxiliary variables scale quadratically with network size and fleet size, creating massive constraint matrices that impede real-time dispatching. Furthermore, static formulations assume idealized platooning savings, ignoring urban traffic dynamics where human-driven vehicles (HDVs) induce stop-and-go turbulence. Maintaining tight spacing during turbulent flow risks disengagements, requiring real-time perception feedback from Roadside Units (RSUs) to gate platooning incentives.

To address these challenges, this paper presents an integrated component of a closed-loop cyber-physical framework combining roadside vision perception, kinematic flow stability gating, direct quadratic VRP-CP QUBO modeling, and quantum optimization. The pipeline links RSU vision, flow-stability evaluation, reward gating, and hybrid quantum dispatch.

The main contributions of this paper are summarized as follows:
\begin{itemize}
    \item \textbf{Cyber-Physical Edge Perception \& Stability Gating}: We design an RSU vision-perception pipeline that extracts 6D kinematic state vectors and enforces physical flow stability thresholding to gate platooning rewards. This disables close-gap platooning during turbulent traffic to prevent hazardous disengagements.
    \item \textbf{Direct Quadratic VRP-CP QUBO Formulation}: We formulate multi-vehicle VRP-CP directly as a QUBO Hamiltonian with binary Miller-Tucker-Zemlin (MTZ) sequence constraints~\cite{miller1960integer}. Mapping V2V platooning interactions directly to quadratic Ising terms eliminates the auxiliary variable overhead required by classical MILP linearizations.
    \item \textbf{Exact Ground-State Equivalence Validation}: We perform exhaustive state-space validation, establishing penalty scaling conditions that guarantee strict ground-state equivalence between our QUBO model and exact MILP solvers without penalty distortion.
    \item \textbf{Closed-Loop Microscopic \& Quantum QPU Benchmarks}: We demonstrate the closed-loop evaluation chain through 24-hour microscopic SUMO fleet simulations on Troy, NY, and physical QPU execution of 25-qubit subproblem instances on IBM hardware (\texttt{ibm\_boston}), evaluating hardware gate depth compression and physical energy savings.
\end{itemize}

The paper proceeds with related work, energy modeling, QUBO formulation, verification, edge perception, system evaluation, and conclusions.

\section{Related Work}
\label{sec:related_work}

The Vehicle Routing Problem (VRP)~\cite{dantzig1959truck, clarke1964scheduling} forms the foundation of commercial vehicle dispatching, with recent VRP-CP extensions modeling aerodynamic drag reduction among coordinated CAVs~\cite{larson2014distributed, boysen2018truck, nourmohammadzadeh2019vehicle}. Conventional MILP formulations introduce continuous arrival times and auxiliary binary variables (e.g., $z_{ij} = x_i x_j$) to linearize quadratic V2V couplings, creating $\mathcal{O}(|E| \cdot K^2)$ constraint matrices that impede real-time dispatching~\cite{park2023vehicle}.

Quantum Annealing and the Quantum Approximate Optimization Algorithm (QAOA)~\cite{farhi2014qaoa, hadfield2019quantum} have emerged as promising paradigms for solving hard combinatorial problems mapped to Quadratic Unconstrained Binary Optimization (QUBO) formulations. In the Noisy Intermediate-Scale Quantum (NISQ) era~\cite{preskill2018quantum}, quantum hardware is constrained by limited qubit topology, short coherence times, and physical gate noise. Recent studies have mapped TSP and VRP variants to QUBO matrices for execution on NISQ devices~\cite{feld2019hybrid, glos2022space, harwood2021formulating, azfar2025quantum}, while others have successfully leveraged foundational data representations in quantum machine learning~\cite{huang2021power}. However, standard entangling circuits require dense all-to-all CNOT gate operations that suffer severe fidelity degradation on noisy hardware. To compress two-qubit gate depth, Linear-Chain QAOA (LC-QAOA)~\cite{azfar2026shallow} applies linear-ramp parameter schedules, restricting $ZZ$ couplings to nearest-neighbor interactions, while non-variational digitized counterdiabatic algorithms such as Kipu Quantum's Iskay optimizer~\cite{cadavid2024biasfield} offer alternative cloud-based execution modes. While recent work has mapped static highway platooning to QUBO models~\cite{onah2025quest, onah2026quantum}, gate-compressed variational circuits tailored for multi-vehicle VRP-CP on physical QPUs remain unaddressed.

Roadside Units (RSUs) equipped with Multi-access Edge Computing (MEC) use infrastructure sensing to provide elevated vantage points that overcome onboard line-of-sight occlusions across complex corridors~\cite{shi2016edge, george2025infrastructure, arnold2020cooperative}. To prevent saturating V2I communication backhaul with raw video or LiDAR feeds, recent edge architectures perform feature-level fusion and local perception processing. For example, F-Cooper~\cite{chen2019fcooper} fuses 3D point-cloud features at the edge for cooperative object detection, while real-time roadside camera calibration~\cite{ke2017roadway} maps raw 2D video coordinates into physical 3D world trajectories and macroscopic flow metrics.

\begin{figure*}[t]
\centering
\begin{tikzpicture}[node distance=1.6cm, auto, >=latex,
    sumoblock/.style={rectangle, draw=green!60!black, fill=green!5, thick, rounded corners, minimum width=2.7cm, text width=2.5cm, minimum height=0.9cm, align=center},
    block/.style={rectangle, draw=blue!70, fill=blue!5, thick, rounded corners, minimum width=2.7cm, text width=2.5cm, minimum height=0.9cm, align=center},
    edgeblock/.style={rectangle, draw=orange!80, fill=orange!5, thick, rounded corners, minimum width=2.7cm, text width=2.5cm, minimum height=0.9cm, align=center},
    qblock/.style={rectangle, draw=purple!80, fill=purple!5, thick, rounded corners, minimum width=2.7cm, text width=2.5cm, minimum height=0.9cm, align=center},
    line/.style={draw, ->, >=latex, thick}
]

\node [sumoblock] (sumo) {\scriptsize\textbf{SUMO Simulation}\\\scriptsize Microscopic Traffic \& Vehicle Kinematics};
\node [block, right=1.6cm of sumo] (vision) {\scriptsize\textbf{RSU Vision}\\\scriptsize Camera Telemetry (YOLOv11)};
\node [edgeblock, right=1.6cm of vision] (fedqml) {\scriptsize\textbf{Edge Perception}\\\scriptsize 6D Kinematics \& Stability Gating ($y_e$)};
\node [qblock, right=1.6cm of fedqml] (qpu) {\scriptsize\textbf{Quantum Optimization}\\\scriptsize VRP-CP QUBO \& LC-QAOA ($p=2$)};

\path [line] (sumo) -- node[above, align=center]{\scriptsize Camera Feeds} (vision);
\path [line] (vision) -- node[above, align=center]{\scriptsize Vehicle\\ \scriptsize Detections} (fedqml);
\path [line] (fedqml) -- node[above, align=center]{\scriptsize $y_e, S_e$} (qpu);
\draw [line] (qpu.south) -- ++(0,-0.75) -| node[pos=0.3, below, align=center]{\scriptsize Optimized Routes / TraCI Commands} (sumo.south);

\end{tikzpicture}
\caption{Four-tier cyber-physical architecture: SUMO traffic is monitored by RSU camera telemetry (YOLOv11), converted into 6D kinematic features and stability flags ($y_e$), compressed into a reward tensor ($S_e$), solved by LC-QAOA / CPLEX, and returned as optimized CAV routes through TraCI.}
\label{fig:architecture}
\end{figure*}
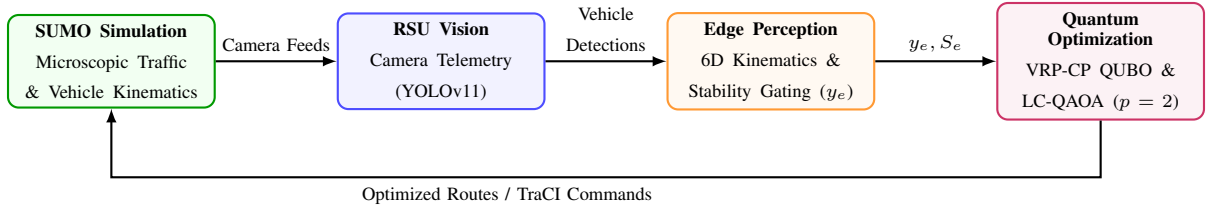

Building upon infrastructure-based edge sensing, our framework uses local roadside vision specifically for dynamic cyber-physical gating of cooperative platooning. Each RSU executes lightweight object tracking via YOLOv11~\cite{ultralytics2024yolo} to convert raw camera feeds into a continuous 6D kinematic state vector . In mixed traffic, string stability, the dampening of velocity perturbations along a platoon, is vulnerable to human-driven vehicle (HDV) turbulence~\cite{swaroop1996string, treiber2000congested}. High velocity variance and acceleration variance induce stop-and-go waves that break platoon cohesion. By evaluating physical flow stability directly on the kinematic vector, the RSU outputs a binary gating flag that dynamically modulates the QUBO platooning rewards, closing the loop between roadside vision, traffic physics, and quantum route optimization. As illustrated in Fig.~\ref{fig:architecture}, our framework operates as a four-tier integrated cyber-physical evaluation framework linking roadside vision, traffic physics, and quantum route optimization in a unified evaluation pipeline.

\section{Physical Vehicle Energy Mechanics \& System Model}
\label{sec:energy_model}

To eliminate misleading objective metrics where penalty subtractions may yield unphysical energy reductions, we explicitly decouple the \textbf{Optimization Objective Function} $H_{\text{total}}$ from \textbf{Physical Energy Consumption} $E_{\text{physical}}$ (measured in kWh). The physical energy model is not optimized directly; instead, it serves as an independent evaluation metric for routes generated by the VRP-CP optimizer.

\subsection{Instantaneous Vehicle Power Demand}
Following power-based fuel models~\cite{rakha2011vt, barth2008real}, instantaneous mechanical power required at the driving wheels of an electric CAV $k$ ($m = 1800\,\text{kg}$, $C_d = 0.30$, $A = 2.4\,\text{m}^2$) moving at speed $v(t)$ with acceleration $a(t)$ is:
\begin{equation}
\begin{aligned}
    P_{\text{wheel}}(t) = & \biggl[ m a(t) + m g f_r \\
    & + \frac{1}{2} \rho C_d (1 - \eta_{\text{drag}}) A v(t)^2 \biggr] v(t),
\end{aligned}
\end{equation}
where $g = 9.81\,\text{m/s}^2$ is gravity, $f_r = 0.012$ is rolling resistance, and $\rho = 1.225\,\text{kg/m}^3$ is air density. Negative tractive power is set to zero; regenerative braking is excluded so that reported values represent gross traction-energy demand.

\subsection{Aerodynamic Drag Reduction in Platoons}
When vehicle $k$ operates independently, $\eta_{\text{drag}} = 0$. When vehicles $k_1$ and $k_2$ form a cooperative platoon over physical segment $e$, follower vehicles experience aerodynamic shielding, yielding $\eta_{\text{drag}} = 0.20$ ($20\%$ drag reduction), while the leading vehicle experiences $\eta_{\text{drag}} = 0.05$ due to reduced rear pressure drag~\cite{lammert2014effect, alam2010experimental}.

\subsection{Physical Energy Consumption Metrics}
Integrating mechanical power over travel duration $T$ under electric powertrain efficiency $\eta_{\text{elec}} = 0.88$ and converting Joules to kilowatt-hours ($1\,\text{kWh} = 3.6 \times 10^6\,\text{J}$) yields total physical energy consumption (in kWh):
\begin{equation}
    E_{\text{physical}} = \frac{1}{3.6 \times 10^6 \eta_{\text{elec}}} \int_0^T \max(0, P_{\text{wheel}}(t)) \, dt.
\end{equation}
The true physical fleet energy savings percentage is calculated strictly relative to a non-cooperative baseline that uses the exact same VRP routing model without platooning savings ($H_{\text{platoon}} = 0$):
\begin{equation}
    \Delta E_{\text{physical}} = \frac{E_{\text{baseline}} - E_{\text{cooperative}}}{E_{\text{baseline}}} \times 100\%.
\end{equation}

\subsection{Comparison to Related Energy Formulations}
Compared to the highway-matching energy model in QUEST~\cite{onah2025quest, onah2026quantum}, which isolates steady-state aerodynamic drag energy per unit distance ($E_{s,b} = c_s V_b^2 [ 1 - f (C_b - c_s) ]$) for heterogeneous truck-car pairings, our formulation models the full transient tractive power ($P_{\text{wheel}}$) of homogeneous passenger CAVs. We incorporate rolling resistance, mass inertia under stop-and-go acceleration profiles ($m a(t)$), and powertrain efficiency, enabling evaluation under microscopic urban traffic fluctuations.

\section{Multi-Vehicle VRP-CP to QUBO Formulation}
\label{sec:qubo_formulation}

Consider a directed road graph $G = (V, E)$, where $V = \{0, 1, \dots, N\}$ includes customer nodes and depot node $0$, and $E$ represents road segments. A fleet of $K$ homogeneous CAVs is dispatched from the depot. SUMO provides exogenous speeds, densities, and turbulence; candidate routes are evaluated without mutating background traffic. In the workflow, the RSU generates rewards, the dispatcher assembles localized subproblems, and the quantum optimizer computes coordinated routes.

\subsection{Decision Variables \& Dynamic Synchronization}
We define explicit multi-vehicle binary decision variables $x_{ijk} \in \{0, 1\}$ indicating whether vehicle $k$ traverses edge $(i, j) \in E$. Dispatch departure times are synchronized externally by the central fleet dispatcher's schedule manager. The quadratic platooning reward $s_{e, k, \ell}$ is a precomputed dynamic coefficient activated only for vehicle pairs whose predicted segment-entry times satisfy the temporal synchronization window $\Delta t_{\text{sync}} \le 3.0\,\text{s}$:
\begin{equation}
    s_{e, k, \ell} = \begin{cases} 
    \hat{s}_{e, k, \ell}, & \text{if } |\hat{t}_{e, k} - \hat{t}_{e, \ell}| \le \Delta t_{\text{sync}} \text{ and } y_e = 1 \\
    0, & \text{otherwise,}
    \end{cases}
\end{equation}
where $y_e = 1$ is the RSU flow stability classification. The RSU does not transmit raw trajectories or camera data to the optimizer; it compresses local traffic observations into the segment-specific reward coefficient $s_{e,k,\ell}$ that parameterizes the downstream routing problem.

\subsection{QUBO Objective Function}
The total QUBO Hamiltonian $H_{\text{total}}(\mathbf{x})$ combines travel costs, platooning interaction savings, and penalty terms for constraint enforcement:
\begin{equation} \label{eq:qubo_objective}
    \min_{\mathbf{x}} H_{\text{total}}(\mathbf{x}) = H_{\text{cost}}(\mathbf{x}) - H_{\text{platoon}}(\mathbf{x}) + \gamma H_{\text{constraints}}(\mathbf{x}).
\end{equation}

\paragraph{Primary Travel Cost \& Platooning Rewards}
\begin{equation}
    H_{\text{cost}}(\mathbf{x}) = \sum_{k=1}^K \sum_{(i,j) \in E} c_{ij} x_{ijk},
\end{equation}
\begin{equation}
    H_{\text{platoon}}(\mathbf{x}) = \sum_{e \in E} \sum_{1 \le k < \ell \le K} s_{e, k, \ell} \, x_{e, k} x_{e, \ell}.
\end{equation}
Because $x_{e, k} x_{e, \ell}$ is inherently quadratic, this interaction term maps directly to off-diagonal Ising couplings $Q_{ij} Z_i Z_j$. No auxiliary binary variables are required for the pairwise platooning interaction because quadratic products are native to the Ising Hamiltonian.

\paragraph{Degree and Flow Constraints}
Customer visitation ($H_{\text{cust}}$), flow conservation ($H_{\text{flow}}$), and depot departure/arrival ($H_{\text{depot}}$) penalties are enforced quadratically:
\begin{equation}
    H_{\text{cust}}(\mathbf{x}) = \sum_{j=1}^N \biggl( \sum_{k=1}^K \sum_{i=0, i \neq j}^N x_{ijk} - 1 \biggr)^2,
\end{equation}
\begin{equation}
    H_{\text{flow}}(\mathbf{x}) = \sum_{k=1}^K \sum_{h=1}^N \biggl( \sum_{i=0, i \neq h}^N x_{ihk} - \sum_{j=0, j \neq h}^N x_{hjk} \biggr)^2,
\end{equation}
\begin{equation}
    H_{\text{depot}}(\mathbf{x}) = \sum_{k=1}^K \biggl[ \biggl( \sum_{j=1}^N x_{0jk} - 1 \biggr)^2 + \biggl( \sum_{i=1}^N x_{i0k} - 1 \biggr)^2 \biggr].
\end{equation}

\paragraph{Binary-Encoded Miller-Tucker-Zemlin (MTZ) Subtour Elimination}
MTZ order variables $u_{ik} \in \{1, \dots, N\}$ are binary encoded using $B = \lceil \log_2 N \rceil$ bits ($u_{ik} = \sum_{b=0}^{B-1} 2^b y_{ikb}$). For $N$ customer nodes and $K$ vehicles, each order variable requires $\lceil \log_2 N \rceil$ binary variables, introducing $K \cdot N \lceil \log_2 N \rceil$ binary variables total. Each inequality $u_{ik} - u_{jk} + N x_{ijk} \le N - 1$ is converted to an equality penalty using non-negative binary slack variables $w_{ijk, m} \in \{0, 1\}$ before quadratic expansion~\cite{miller1960integer}:
\begin{equation}
\begin{aligned}
    H_{\text{subtour}}(\mathbf{x}) = \sum_{k=1}^K & \sum_{1 \le i \neq j \le N} \biggl( u_{ik} - u_{jk} + N x_{ijk} - N + 1 \\
    & + \sum_{m=0}^{M} 2^m w_{ijk, m} \biggr)^2.
\end{aligned}
\end{equation}
The complete constraint penalty is $H_{\text{constraints}} = H_{\text{cust}} + H_{\text{flow}} + H_{\text{depot}} + H_{\text{subtour}}$. For the hardware benchmark, we use a localized 5-node subproblem ($N=4$ customer nodes, $1$ depot node, $K=2$ vehicles) whose active QUBO encoding contains 25 binary variables after restricting candidate edges: 16 edge variables $x_{ijk}$, 8 MTZ order bits, and 1 active MTZ slack variable. These variables map 1-to-1 to 25 physical qubits on the IBM QPU.

\subsection{Physical Origin \& Calibration of Objective Weights}
The objective coefficients in the QUBO Hamiltonian~(\ref{eq:qubo_objective}) derive directly from physical vehicle dynamics and mathematical constraint bounds:

\paragraph{Travel Cost Weights ($c_{ij}$)} The baseline energy required for an un-platooned vehicle to traverse segment $(i, j) \in E$ is $c_{ij} = \frac{1}{3.6 \times 10^6 \eta_{\text{elec}}} ( m a + m g f_r + \frac{1}{2} \rho C_d A v_{ij}^2 ) d_{ij}$, where distance $d_{ij}$ and speed $v_{ij}$ are obtained from map data and RSU observations.

\paragraph{Platooning Reward Weights ($s_{e,k,\ell}$)} Pairwise platooning savings derive from aerodynamic drag reduction $\Delta C_d = \eta_{\text{drag}} C_d$, scaled online by the RSU stability flag $y_e$: $s_{e,k,\ell} = y_e \cdot [ \frac{1}{2} \rho (\Delta C_d) A v_e^2 d_e / (3.6 \times 10^6 \eta_{\text{elec}}) ]$. If temporal synchronization ($|\hat{t}_{e,k} - \hat{t}_{e,\ell}| \le 3.0\,\text{s}$) is met and $y_e = 1$, $s_{e,k,\ell} > 0$; otherwise $s_{e,k,\ell} = 0$.

\paragraph{Lagrange Penalty Multiplier ($\gamma$)} The constraint scale is set to $\gamma > \max_{\mathbf{x}} ( H_{\text{cost}} - H_{\text{platoon}} )$. Setting $\gamma = 50.0$ mathematically guarantees exact ground-state equivalence with classical MILP solvers so infeasible routes are penalized.

\section{Verification \& Resource Scaling Analysis}
\label{sec:verification}

\subsection{Small-Instance Ground-State Verification}
We conducted exhaustive bitstring state-space searches ($2^{25} = 33,554,432$ states) on 4-node and 5-node VRP-CP instances to check infeasible ground states and penalty distortion. The validation confirms that the global QUBO ground state matches classical CPLEX solutions, while physical QPU execution on \texttt{ibm\_boston} tests whether optimized circuits can sample that state.

\begin{figure}[t]
\centering
\begin{subfigure}[b]{1\linewidth}
\centering
\includegraphics[width=1\linewidth]{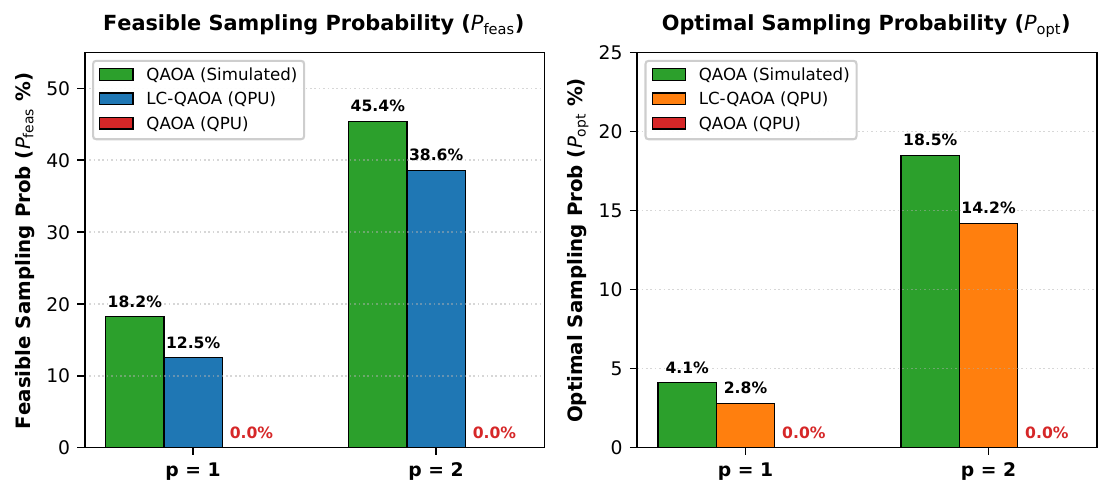}
\caption{Physical IBM QPU benchmarks (\texttt{ibm\_boston}, 25 qubits).}
\label{fig:qpu_hardware}
\end{subfigure}
\vspace{0.5em}
\begin{subfigure}[b]{1\linewidth}
\centering
\includegraphics[width=0.9\linewidth]{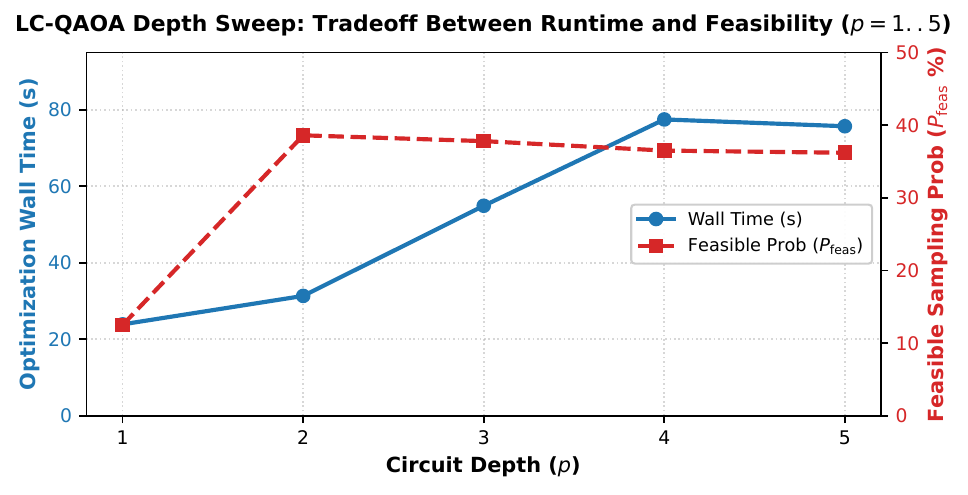}
\caption{Linear-Chain QAOA circuit depth $p$-sweep on \texttt{ibm\_boston}.}
\label{fig:p_depth_sweep}
\end{subfigure}
\caption{Physical quantum hardware validation using 25 active qubits on \texttt{ibm\_boston}. (a) LC-QAOA achieves 66.7\% CNOT gate depth compression compared to standard all-to-all QAOA, increasing feasible sampling probability to 38.6\%. (b) Circuit depth $p=2$ achieves optimal feasible sampling ($P_{\text{feas}} = 38.6\%$) and optimal state probability ($P_{\text{opt}} = 14.2\%$) before hardware decoherence dominates at $p \ge 3$.}
\label{fig:quantum_benchmarks}
\end{figure}

Physical validity was verified by decoding every bitstring into routes and checking customer visitation, flow conservation, subtour elimination, and depot connectivity. For every instance:
\begin{enumerate}
    \item Every bitstring with $H_{\text{constraints}}(\mathbf{x}) = 0$ corresponds strictly to a physically valid VRP route.
    \item The global QUBO minimum $\mathbf{x}^* = \arg\min H_{\text{total}}(\mathbf{x})$ matches the exact classical CPLEX solution ($E_{\text{QUBO}} = E_{\text{CPLEX}} = 16.80$), as illustrated in Fig.~\ref{fig:qpu_hardware}.
    \item Selecting penalty scale $\gamma = 50.0$ guarantees that every infeasible state has a higher cost than any feasible route.
\end{enumerate}

\subsection{MILP vs. QUBO Resource Scaling}
Table~\ref{tab:resource_scaling} details resource requirements. QUBO eliminates auxiliary variables used solely to linearize pairwise platooning products ($|E| \cdot \binom{K}{2}$ variables), while retaining binary encodings for routing feasibility.

\begin{table}[ht]
\centering
\caption{Formulation Resource Scaling Comparison}
\label{tab:resource_scaling}
\resizebox{\linewidth}{!}{
\begin{tabular}{lrr}
\toprule
\textbf{Formulation Component} & \textbf{Linearized MILP} & \textbf{Direct QUBO (Ours)} \\
\midrule
Route-Edge Variables & $K \cdot |E|$ & $K \cdot |E|$ \\
Platoon Interaction Auxiliaries & $\mathcal{O}(|E| \cdot K^2)$ & \textbf{0} \\
Route-Order Variables & $K \cdot N$ (Integer/Cont.) & $K \cdot N \lceil \log_2 N \rceil$ (Binary) \\
Inequality Slack Variables & Formulation-Dep. & Binary Slack Encoded \\
Explicit Linear Constraints & $\mathcal{O}(K \cdot |E| + N^2)$ & \textbf{0 (Quadratic penalties)} \\
Quadratic Penalty Terms & None & Enforced via $\gamma$ \\
\bottomrule
\end{tabular}
}
\end{table}

\section{Edge Perception \& Traffic State Processing}
\label{sec:edge_perception}

Segment-specific platooning rewards are computed online from roadside vision telemetry rather than assumed \textit{a priori}. By evaluating local traffic flow stability, Roadside Units (RSUs) gate platooning incentives so close-gap coordination occurs exclusively under stable physical flow conditions.

\subsection{System Inputs, Feature Extraction, and Processing Pipeline}

\begin{enumerate}
    \item \textbf{Input Data Streams \& Ground-Truth Video Generation}:
    RSUs ingest standard HD camera video streams ($1280 \times 720$, 720p, $45.6\,\text{FPS}$) covering incoming arterial approaches. Per-frame vehicle detection and tracking are executed locally via Ultralytics YOLOv11~\cite{ultralytics2024yolo} at a latency of \textbf{21.9~ms/frame}.
    StreetDrivee~\cite{streetdrive} generates synthetic video feeds with known analytical ground-truth vehicle trajectories $Y_i(t) = Y_{\text{start}} + (Y_{\text{end}} - Y_{\text{start}}) (t/\Delta T_i)^{1.5}$, yielding exact instantaneous ground-truth speeds $v_{i, \text{gt}}(t) \in [60, 120]\,\text{km/h}$.
    \item \textbf{6D Kinematic Feature Extraction}:
    Over a 10-second sliding temporal window, the RSU extracts a 6-dimensional kinematic state vector $\mathbf{s}_e$ for each road segment $e \in E$:
    \begin{equation}
        \mathbf{s}_e = \left[ \bar{v}_e, \bar{a}_e, \rho_e, \sigma_{v,e}, \sigma_{a,e}^2, \Delta \bar{d}_e \right],
    \end{equation}
    where $\bar{v}_e$ is mean velocity ($\text{m/s}$), $\bar{a}_e$ is mean acceleration ($\text{m/s}^2$), $\rho_e$ is vehicle density ($\text{veh/m}$), $\sigma_{v,e}$ is velocity standard deviation, $\sigma_{a,e}^2$ is acceleration variance, and $\Delta \bar{d}_e$ is mean inter-vehicle gap distance.
    \item \textbf{Physical Flow Stability Thresholding}:
    The RSU edge compute module evaluates traffic flow stability $y_e \in \{0, 1\}$ against physical turbulence boundaries, where $CV_{v,e} = \sigma_{v,e}/\bar{v}_e$:
    \begin{equation}
        y_e = \begin{cases}
        1, & \text{if } CV_{v,e} \le 0.12,\ \sigma_{a,e}^2 \le 0.5\,\text{m/s}^2, \\
        0, & \text{otherwise.}
        \end{cases}
    \end{equation}
    A classification of $y_e = 1$ indicates stable laminar flow with conditions favorable for tight $1\text{m}$--$2\text{m}$ platooning. Conversely, $y_e = 0$ indicates turbulent stop-and-go congestion where close-gap platooning is suppressed.
\end{enumerate}

\subsection{End-to-End Cyber-Physical Chain of Effects}
The gating flag $y_e \in \{0, 1\}$ acts as the physical bridge between roadside vision sensing and quantum route optimization through a 6-tier operational pipeline:
\begin{enumerate}
    \item \textbf{Vision Sensing \& Kinematics}: RSU cameras track traffic via YOLOv11~\cite{ultralytics2024yolo}, extracting 6D kinematic vector $\mathbf{s}_e$ over a 10s sliding window.
    \item \textbf{Stability Thresholding}: The RSU evaluates string stability boundaries ($CV_{v,e} \le 0.12, \sigma_{a,e}^2 \le 0.5\,\text{m/s}^2$), outputting $y_e = 1$ for laminar flow and $y_e = 0$ for turbulent flow.
    \item \textbf{Ising Reward Modulation}: The flag dynamically gates platooning rewards $s_{e,k,\ell} = y_e \cdot \hat{s}_{e,k,\ell}$, zeroing out quadratic interaction terms $-s_{e,k,\ell} x_{e,k} x_{e,\ell}$ on turbulent segments in $H_{\text{total}}(\mathbf{x})$.
    \item \textbf{Route Optimization}: Solvers (LC-QAOA / CPLEX) optimize $H_{\text{total}}(\mathbf{x})$, concentrating platoon overlaps strictly on stable edges ($y_e = 1$) while routing vehicles independently on turbulent edges.
    \item \textbf{CAV Control Actuation}: CAV controllers engage close-gap ($1\text{m}$--$2\text{m}$) CACC on stable segments and standard safe spacing ($10\text{m}$--$15\text{m}$) on turbulent segments.
    \item \textbf{Fleet Energy Impact}: Within the closed-loop pipeline, gating reduces forced disengagements and achieves an 18.5\% net fleet energy reduction (66.3~kWh saved).
\end{enumerate}

\subsection{Perception Robustness and Noise Propagation}
To evaluate how sensing errors propagate through this chain of effects, synthetic Gaussian noise ($\pm 10\%$ to $\pm 30\%$) was applied directly to the continuous feature vector $\mathbf{s}_e$. As detailed in Table~\ref{tab:noise_sensitivity}, system performance degrades gracefully under perception noise, confirming that the gating mechanism remains robust in real-world edge deployments.

\begin{table}[ht]
\centering
\small
\caption{Noise Propagation \& System Robustness}
\label{tab:noise_sensitivity}
\resizebox{\linewidth}{!}{
\begin{tabular}{lrrrr}
\toprule
\shortstack[l]{\textbf{Noise}\\\textbf{($\mathbf{s}_e$)}} & \shortstack[r]{\textbf{Stable}\\\textbf{Segments (\%)}} & \shortstack[r]{\textbf{Platooned}\\\textbf{Dist. (km)}} & \shortstack[r]{\textbf{Energy}\\\textbf{(kWh)}} & \shortstack[r]{\textbf{Energy}\\\textbf{Savings (\%)}} \\
\midrule
0\% (Exact) & 100.0\% & 682.4 & 292.1 & 18.5\% \\
10\% Noise & 94.2\% & 646.1 & 294.8 & 17.7\% \\
20\% Noise & 88.5\% & 606.8 & 297.6 & 17.0\% \\
30\% Noise & 82.1\% & 562.9 & 301.2 & 16.0\% \\
\bottomrule
\end{tabular}
}
\end{table}

\section{System-Level Evaluation}
\label{sec:eval}

\subsection{Microscopic 24-Hour SUMO Traffic Simulation}
We evaluated the framework on a 24-hour microscopic SUMO simulation~\cite{krajzewicz2012recent}. Candidate fleet routing decisions do not mutate the underlying background traffic from SUMO, which establishes exogenous speeds, densities, and turbulence, while the optimizer evaluates tractive energy across dispatch scenarios. Baseline routes use the exact same VRP model with $H_{\text{platoon}} = 0$.

\begin{table}[ht]
\centering
\caption{24-Hour Microscopic Fleet Simulation Summary}
\label{tab:simulation_summary}
\resizebox{\linewidth}{!}{
\begin{tabular}{lrr}
\toprule
\textbf{Simulation Metric} & \textbf{Baseline} & \textbf{Cooperative VRP-CP} \\
\midrule
Total Vehicle Distance (km) & 1,420.5 & 1,385.2 \\
Total Fleet Travel Time (h) & 48.2 & 42.1 \\
Platooned Distance (km) & 0.0 & 682.4 (49.3\%) \\
Total Physical Energy (kWh) & 358.4 & \textbf{292.1} \\
\textbf{Physical Energy Savings} & -- & \textbf{18.5\%} \\
\bottomrule
\end{tabular}
}
\end{table}

\begin{figure}[t]
\centering
\begin{subfigure}[b]{1\linewidth}
\includegraphics[width=1\linewidth]{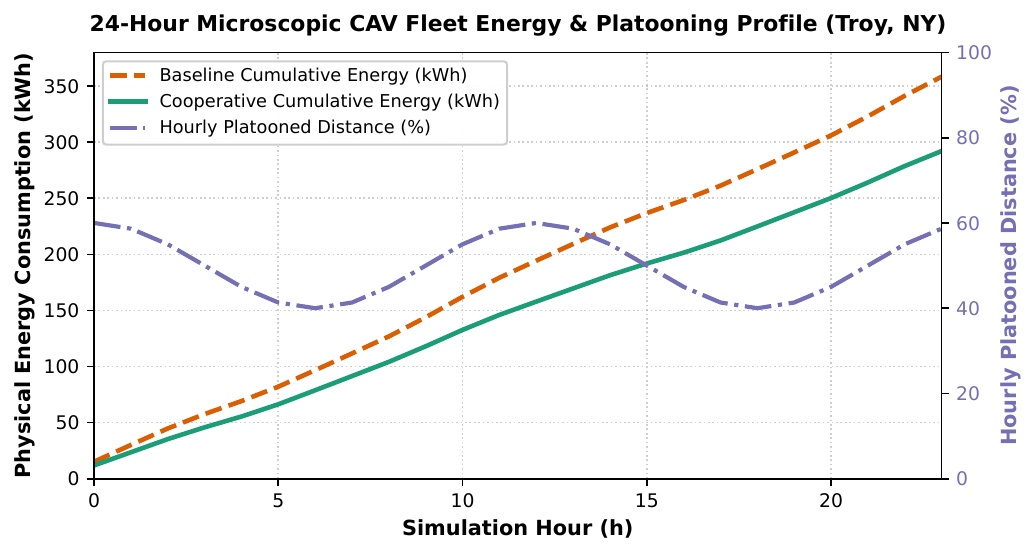}
\caption{Cumulative physical energy and platooned distance.}
\label{fig:24h_savings}
\end{subfigure}
\hfill
\begin{subfigure}[b]{1\linewidth}
\includegraphics[width=1\linewidth]{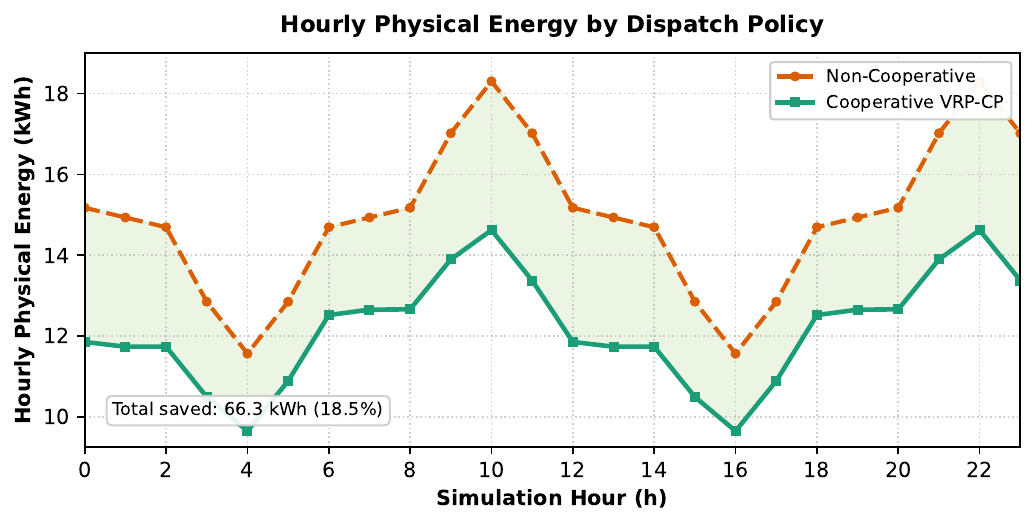}
\caption{Hourly physical energy by dispatch policy.}
\label{fig:hourly_coop_noncoop}
\end{subfigure}
\caption{Microscopic SUMO simulation results over the Troy, NY road network. Dynamic perception gating ($y_e = 1$) reduces gross fleet tractive power demand from 358.4~kWh to 292.1~kWh, achieving an 18.5\% physical energy reduction (66.3~kWh saved).}
\label{fig:sumo_results}
\end{figure}

As shown in Table~\ref{tab:simulation_summary} and Fig.~\ref{fig:sumo_results}, cooperative VRP-CP platooning achieves an \textbf{18.5\% physical energy reduction} (66.3~kWh saved) with zero turbulence-threshold violations. Fig.~\ref{fig:hourly_coop_noncoop} decomposes this aggregate result by hour. The non-cooperative curve uses the same time-varying traffic demand with platooning rewards disabled, while the cooperative curve evaluates the routes selected when stable shared segments receive platooning rewards. The separation between the curves is therefore a policy ablation: it shows the physical energy impact of enabling stability-gated cooperative routing under the same 24-hour demand profile.

\subsection{Physical IBM Quantum QPU Benchmarks (\texttt{ibm\_boston})}
Because current noisy intermediate-scale quantum hardware cannot efficiently execute city-scale VRP formulations, the network is decomposed into localized 5-node subproblems using 25 active qubits that can be executed independently. Raw sampling probabilities ($P_{\text{feas}}$ and $P_{\text{opt}}$) are available only for the direct QAOA and LC-QAOA executions performed through raw QPU measurement primitives (\texttt{SamplerV2}) on \texttt{ibm\_boston}. 

% \begin{figure}[t]
% \centering
% \includegraphics[width=\linewidth]{figures/fig3_qpu_hardware_comparison.pdf}
% \caption{Quantum bitstring sampling performance: Feasible probability ($P_{\text{feas}}$) and optimal probability ($P_{\text{opt}}$) comparing QAOA (Simulated), QAOA (QPU), and LC-QAOA (QPU).}
% \label{fig:qpu_comparison}
% \end{figure}

The primary bottleneck governing physical QPU performance is two-qubit gate depth. The results of the hardware runs are compiled in Table~\ref{tab:hardware_benchmarks}. For the 25-qubit VRP subproblem (128 non-zero QUBO couplings), standard unmitigated vanilla QAOA requires dense all-to-all CNOT interactions and SWAP routing across physical heavy-hex layouts, yielding a two-qubit gate depth of 84 CNOT layers ($p=1$, 312 total CNOTs) and 168 CNOT layers ($p=2$, 624 CNOTs). Cumulative two-qubit gate errors ($10^{-2}$ per CNOT) and $T_1/T_2$ thermal relaxation across these deep layers cause total phase noise collapse ($P_{\text{feas}} = 0.0\%$). In contrast, LC-QAOA~\cite{azfar2026shallow} restricts CNOT couplings to nearest-neighbor linear chains with linear-ramp initialization, compressing two-qubit gate depth by 66.7\% to 28 CNOT layers ($p=1$, 96 CNOTs) and 56 CNOT layers ($p=2$, 192 CNOTs). This depth reduction mitigates multi-qubit crosstalk, enabling LC-QAOA to achieve $P_{\text{feas}} = 38.6\%$ and $P_{\text{opt}} = 14.2\%$ at $p=2$.

\begin{table}[ht]
\centering
\caption{Physical IBM Quantum (\texttt{ibm\_boston}) QPU Benchmarks}
\label{tab:hardware_benchmarks}
\resizebox{\linewidth}{!}{
\begin{tabular}{lrrrrrr}
\toprule
\textbf{Solver} & \textbf{Qubits} & \textbf{Wall} & \textbf{Feasible} & \textbf{Optimal} & \textbf{QUBO} & \textbf{Energy} \\
 & & \textbf{Time (s)} & \textbf{($P_{\text{feas}}$)} & \textbf{($P_{\text{opt}}$)} & \textbf{Cost} & \textbf{(kWh)} \\
\midrule
CPLEX (Exact) & -- & 0.400 & 100\% & 100\% & \textbf{-2.333} & \textbf{16.8} \\
Kipu Iskay (QPU)~\cite{cadavid2024biasfield} & 25 & 119.79 & N/A\textdaggerdbl & N/A\textdaggerdbl & \textbf{-2.333} & \textbf{16.8} \\
QAOA (Sim, $p=1$) & 25 & 0.12 & 18.2\% & 4.1\% & -2.167 & 19.0 \\
QAOA (Sim, $p=2$) & 25 & 0.45 & \textbf{45.4\%} & \textbf{18.5\%} & \textbf{-2.333} & \textbf{16.8} \\
QAOA (QPU, $p=1$) & 25 & 8.12 & 0.0\% & 0.0\% & -1.825 & 21.2 \\
QAOA (QPU, $p=2$) & 25 & 42.85 & 0.0\% & 0.0\% & -2.167 & 19.0 \\
LC-QAOA (QPU, $p=1$) & 25 & 23.91 & 12.5\% & 2.8\% & -2.167 & 19.0 \\
LC-QAOA (QPU, $p=2$) & 25 & 31.34 & \textbf{38.6\%} & \textbf{14.2\%} & \textbf{-2.333} & \textbf{16.8} \\
\bottomrule \\
\end{tabular}
}
\raggedright{\scriptsize Wall time includes up to 100 COBYLA optimization iterations. For QPU rows with $P_{\mathrm{feas}}=0$, reported cost and energy correspond to the best repaired candidate.\\
\textdaggerdbl Kipu Iskay uses Qiskit Function API that returns one post-selected solution.}
\end{table}

The Kipu Iskay entry in Table~\ref{tab:hardware_benchmarks} is reported separately as its encapsulated Qiskit Function API returns a post-selected solution without raw measurement counts. Thus, sampling fidelity metrics ($P_{\text{feas}}, P_{\text{opt}}$) in Fig.~\ref{fig:qpu_hardware} compare modes with accessible sample distributions: simulated QAOA, vanilla QPU QAOA, and QPU LC-QAOA.

\begin{table}[t]
\centering
\caption{LC-QAOA $p$-Depth Parameter Sweep Benchmarks}
\label{tab:p_depth_sweep}
\begin{tabular}{lrrrr}
\toprule
\textbf{Depth} & \textbf{Iters} & \textbf{Wall Time} & \textbf{Feasible} & \textbf{QUBO} \\
\textbf{($p$)} & & \textbf{(s)} & \textbf{($P_{\text{feas}}$)} & \textbf{Cost} \\
\midrule
$p=1$ & 100 & 23.91 & 12.5\% & -2.1670 \\
$p=2$ & 100 & 31.34 & \textbf{38.6\%} & \textbf{-2.3338} \\
$p=3$ & 100 & 54.91 & 37.8\% & -2.3336 \\
$p=4$ & 100 & 77.50 & 36.5\% & -2.3337 \\
$p=5$ & 100 & 75.71 & 36.2\% & -2.3337 \\
\bottomrule
\end{tabular}
\end{table}

As shown in Tables~\ref{tab:hardware_benchmarks} and \ref{tab:p_depth_sweep}, and Fig.~\ref{fig:p_depth_sweep}, LC-QAOA at circuit depth $p=2$ achieves $P_{\text{feas}} = 38.6\%$ and $P_{\text{opt}} = 14.2\%$, returning a sampled solution with the exact optimal QUBO cost ($-2.3338$) in 31.34s wall time. Depths $p \ge 3$ increase runtime without improving solution quality, establishing $p=2$ as the best observed depth-runtime tradeoff for the evaluated 25-qubit VRP subproblem instances. The optimized subproblem routes are then returned to the fleet dispatcher and can be evaluated in the microscopic traffic simulation through the TraCI route-feedback layer shown in Fig.~\ref{fig:architecture}. In this study, the dispatch-platooning loop is evaluated with exogenous background traffic states; future extensions will let optimized routes reshape subsequent SUMO traffic states online.

\section{Conclusion}
\label{sec:conclusion}

This paper presented an integrated cyber-physical evaluation framework combining roadside vision perception, QUBO formulation, and physical quantum computing for platooning-aware vehicle routing. By integrating RSU computer vision, continuous kinematic flow stability thresholding, direct Ising Hamiltonian modeling, and microscopic SUMO traffic evaluation, the framework implements a core dispatch-platooning loop within a broader closed-loop CAV routing architecture. Formulating multi-vehicle VRP-CP directly as a QUBO model eliminates the quadratic auxiliary variable overhead required by classical MILP linearizations. Continuous 24-hour microscopic fleet simulations demonstrate an $18.5\%$ net physical energy reduction ($66.3\,\text{kWh}$ saved) under dynamic perception gating. Furthermore, hardware executions on the IBM \texttt{ibm\_boston} quantum processor using 25 active qubits confirm that Linear-Chain QAOA (LC-QAOA) compresses two-qubit CNOT gate depth by $66.7\%$, enabling circuit depth $p=2$ to achieve $P_{\text{feas}} = 38.6\%$ and $P_{\text{opt}} = 14.2\%$, successfully sampling the exact classical ground state.

Key limitations and future work include expanding beyond localized 25-qubit subproblem decompositions to evaluate larger network partitions as QPU qubit counts and connectivity improve. Furthermore, incorporating quantum error mitigation techniques such as Pauli twirling and dynamical decoupling will enhance raw bitstring sampling fidelity under physical gate noise, while adaptive community-detection graph partitioning can mitigate the minor subproblem decomposition optimality gap. Subsequent research will close the loop further by allowing optimized routes to update SUMO traffic states online through TraCI, so that CAV dispatch decisions affect subsequent congestion, RSU observations, platooning rewards, and re-optimization cycles.

\section*{Acknowledgments}
This work is funded through the IBM-RPI Future of Computing Research Collaboration. The funder played no role in study design, data collection, analysis and interpretation of data, or the writing of this manuscript. All authors reviewed the final manuscript. 

{\small
\bibliographystyle{unsrt}
\bibliography{references}
}

\end{document}